\documentclass{article}
\usepackage{spconf,amsmath,graphicx}

\usepackage{multirow}
\usepackage{amssymb}
\usepackage{lipsum}
\usepackage{todonotes}
\usepackage{enumitem}

\title{ACOUSTIC MODEL FUSION FOR END-TO-END SPEECH RECOGNITION}
\name{\begin{tabular}{@{}c@{}}
Zhihong Lei, Mingbin Xu, Shiyi Han, Leo Liu, Zhen Huang, Tim Ng, Yuanyuan Zhang, \\Ernest Pusateri, Mirko Hannemann, Yaqiao Deng, Man-Hung Siu
\end{tabular}
}
\address{Apple \\
\texttt{\{zlei, mingbinxu, manhung\_siu\}@apple.com}}

\begin{document}
%
\maketitle
\begin{abstract}
Recent advances in deep learning and automatic speech recognition (ASR) have enabled the end-to-end (E2E) ASR system and boosted the accuracy to a new level. The E2E systems implicitly model all conventional ASR components, such as the acoustic model (AM) and the language model (LM), in a single network trained on audio-text pairs.
Despite this simpler system architecture, fusing a separate LM, trained exclusively on text corpora, into the E2E system 
has proven to be beneficial. However, the application of LM fusion presents certain drawbacks, such as its inability to address the domain mismatch issue inherent to the internal AM.
Drawing inspiration from the concept of LM fusion, we propose the integration of an external AM into the E2E system to better address the domain mismatch. 
By implementing this novel approach, we have achieved a significant reduction in the word error rate, with an impressive drop of up to 14.3\% across varied test sets. We also discovered that this AM fusion approach is particularly beneficial in enhancing named entity recognition.


\end{abstract}
\section{Introduction}
\label{sec:intro}

Deep neural networks have been the dominant modeling technology since their introduction into automatic speech recognition (ASR) \cite{hinton2012deep}. The conventional hybrid ASR systems consist of an acoustic model (AM) and a language model (LM) \cite{dahl2011context}. 
Both can be modeled by sophisticated neural networks.
Despite the superior modeling capability of neural networks, the AM and the LM are separately optimized for the best phonetic accuracy and perplexity (PPL) respectively, which does not necessarily translate to optimal ASR accuracy.

In contrast, end-to-end (E2E) ASR is modeled in a single neural network, predicting a character or word-piece sequence given audio.
AM, LM, and pronunciation dictionary (lexicon) are implicitly learned in the neural network. 
This simplicity also makes training more consistent with the system objective, namely word error rate (WER). The most popular modeling approaches to E2E ASR can be categorized as (a) Connectionist Temporal Classification (CTC) \cite{graves2006connectionist}, (b) Recurrent Neural Network Transducer (RNN-T) \cite{graves2012sequence}, and (c) Attention Encoder Decoder (AED) \cite{cho2014learning, bahdanau2014neural}. 
More recently, self-attention \cite{vaswani2017attention} based models such as Transformer and Conformer \cite{gulati2020conformer} have shown promising results when they are trained with RNN-T loss \cite{gulati2020conformer} or CTC loss \cite{guo2021recent}. 
The essence of end-to-end ASR models requires their training on audio-text pairs. Nevertheless, this characteristic presents a unique hurdle when it comes to rare words that are not adequately represented in paired audio data. Consequently, these words often suffer from poor recognition when compared to the recognition capabilities of hybrid ASR models.

To overcome this, most researches focus on injecting missing/external knowledge into the model by leveraging an external language model via various approaches \cite{kannan2018analysis, gulcehre2015using, sriram2017cold}.
However, these approaches all assume domain invariance of the internal AM, which is unlikely to be true. 
The internal AM performs inadequately on long tail words and phrases as their constituent wordpieces are rarely seen in the training data. 
Moreover, the internal pronunciation lexicon does not well capture the pronunciations of rare or unusually spelled words. 

In this work, we propose to integrate an external AM into the E2E system, called \textbf{AM fusion} hereafter.  
Similar to shallow fusion \cite{kannan2018analysis}, we interpolate an AM score with the E2E system score in log-linear space. 
Our experiments demonstrate that AM fusion is able to
\begin{itemize}[noitemsep]
    \item boost E2E ASR overall accuracy,
    \item inject external knowledge into E2E ASR, and
    \item mitigate the lack of named entity training data and enhance named entity recognition.
\end{itemize}
We report up to 14.3\% relative WER reduction (WERR) on various test sets.

\section{Related Work}
\label{sec:related}

Integrating an external LM has proven to be an effective approach for external knowledge integration with E2E ASR \cite{toshniwal2018comparison}. 
An external LM, typically trained on larger, multi-domain corpora, can effectively enhance underrepresented words and phrases in the E2E training data, and mitigate mismatches between training and testing phases. 
Different approaches are proposed to integrate LMs, such as shallow fusion \cite{kannan2018analysis}, cold fusion \cite{sriram2017cold} and deep fusion \cite{gulcehre2015using}.

In addition to the external LM, researchers argue that there is an internal LM implicitly trained in E2E ASR models. Density Ratio \cite{mcdermott2019density} was brought up to estimate the internal LM by training a neural LM on the transcripts of the E2E training data. The Density Ratio method has demonstrated consistent superiority over shallow fusion, though it does rely heavily on the assumption that E2E models can be factorized into an AM and LM.
The Internal Language Model Estimation (ILME) method \cite{meng2021internal} takes a different approach. It directly approximates the internal LM scores by nullifying the acoustic representation from the network. However, it is worth noting that this straightforward estimation could potentially be overly simplistic and imprecise.



\subsection{Shallow fusion}
\label{subsec:sf}

One of the most popular ways of external language model integration is shallow fusion. It interpolates a separately trained language model with the E2E model in log-linear space. During decoding, a token sequence $ Y  $ that maximizes the joint E2E-LM probabilities is obtained as follows
\begin{equation}
Y = \arg \max_{\hat{Y}}{ \Big[ \log P_{E2E}(\hat{Y} | X) + \lambda
_{LM} \log P_{LM}(\hat{Y}) \Big]}
\label{eq:lmf}
\end{equation}
where $\log P_{E2E}(\hat{Y} | X)$ denotes the log posteriors of the E2E model given the acoustic sequence $X$ and $P_{LM}(\hat{Y})$ denotes the external LM probability of token sequence $\hat{Y}$.  $\lambda_{LM}$ is a weight applied to the external LM. Assume the E2E model can be factorized into an internal AM and an internal LM as follows
\begin{equation}
P_{E2E}(\hat{Y} | X) =  \frac{P_{IAM}(X | \hat{Y}) P_{ILM}(\hat{Y})}  {P(X)}
\label{eq:ilm}
\end{equation}
where $P_{IAM}(X | \hat{Y})$ denotes the internal AM and $P_{ILM}(\hat{Y})$ denotes the internal LM probability.  Combining Eq. (\ref{eq:lmf}) and (\ref{eq:ilm}), we derive an objective function as follows
\begin{multline}
Y = \arg \max_{\hat{Y}}{ \Big[ \log P_{IAM}(X | \hat{Y})} \\ 
 +\log P_{ILM}(\hat{Y}) + \lambda_{LM} \log P_{LM}(\hat{Y}) \Big]
\end{multline}
This is equivalent to a system with the LM being $\log P_{ILM}(\hat{Y})$ \\
$ + \lambda_{LM} \log P_{LM}(\hat{Y})$ and the AM being
$\log P_{IAM}(X | \hat{Y})$.  As the internal LM weight is implicit and untunable, we can only tune the external LM weight to adjust the overall AM/LM combination. This adds some extra LM cost to the system, equivalently diminishing the internal AM.  

\subsection{ILM Negation}
On the other hand, the actual LM is always a combination of the internal and external LM. When the target domain mismatches the source domain,  it is desirable to amplify the influence of the external LM while simultaneously diminishing the contribution of the internal LM.
Assume a target domain distribution as follows
\begin{equation}
P_{{E2E}_t}(\hat{Y} | X) =  \frac{P_{AM}(X | \hat{Y}) P_{LM}(\hat{Y})}  {P(X)}
\end{equation}
where $P_{AM}(X | \hat{Y})$ denotes the target domain's AM. 
Combining Eq.  (2) and (4),  the target domain system can be represented by
\begin{equation}
P_{{E2E}_t}(\hat{Y} | X) = P_{E2E}(\hat{Y} | X) \frac{P_{AM}(X | \hat{Y}) P_{LM}(\hat{Y})}{P_{IAM}(X | \hat{Y}) P_{ILM}(\hat{Y})}.
\end{equation}
With further assumption of domain invariant AM, namely $\log P_{IAM}(X | \hat{Y}) = \log P_{AM}(X | \hat{Y})$, the following objective function is obtained with an internal LM score subtracted. In engineering practices, tunable weights $\lambda_{LM}$ and $\lambda_{ILM}$ are introduced to balance the external and internal language models.
\begin{multline}
Y = \arg \max_{\hat{Y}}{ \Big[ \log P_{E2E}(\hat{Y} | X) +  \lambda_{LM} \log P_{LM}(\hat{Y})} \\
- \lambda_{ILM} \log P_{ILM}(\hat{Y}) \Big].
\end{multline}

\section{Methodology}
\label{sec:method}

\subsection{Acoustic model fusion}

\label{sec:af}

Inspired by prior work of language model fusion, we propose acoustic model fusion. The objective can be described as follows
\begin{multline}
Y = \arg \max_{\hat{Y}}{ \Big[ \log P_{E2E}(\hat{Y} | X) + \lambda
_{AM} \log P_{AM}(X | \hat{Y})} \\ + \lambda_{LM} \log P_{LM}(\hat{Y}) \Big].
\end{multline}
If the E2E model is decomposed into an internal LM and AM, the objective function is then
\begin{multline}
Y = \arg \max_{\hat{Y}}{ \Big[ \log P_{IAM}(X | \hat{Y}) + \lambda
_{AM} \log P_{AM}(X | \hat{Y})} \\
+ \log P_{ILM}(\hat{Y}) + \lambda_{LM} \log P_{LM}(\hat{Y}) \Big].
\label{eq:amf}
\end{multline}
AM fusion provides an equivalent system of an AM 
\begin{equation}
\log P_{IAM}(X | \hat{Y}) + \lambda_{AM} \log P_{AM}(X | \hat{Y})
\end{equation}
and an LM 
\begin{equation}
\log P_{ILM}(\hat{Y}) + \lambda_{LM} \log P_{LM}(\hat{Y}).
\end{equation}
While the external AM can help with under-trained word-pieces with additional pronunciation information, it also introduces an alternative approach to reduction of the influence of the internal LM. If we use the domain invariant assumption of the AM, we can derive the following objective function from Eq. (\ref{eq:amf})
\begin{multline}
Y = \arg \max_{\hat{Y}}{ \Big[ \log P_{IAM}(X | \hat{Y}) + \frac{1}{1+\lambda_{AM}}  \log P_{ILM}(\hat{Y})}\\
 + \frac{\lambda_{LM}}{1+\lambda_{AM}}  \log P_{LM}(\hat{Y}) \Big].
\label{eq:alternative}
\end{multline}
We can manipulate the weights of the external LM ($\lambda_{LM}$) and AM ($\lambda_{AM}$) to achieve any weighted combination of the AM, internal and external LM as per requirement.
However, estimation of AM is usually more complicated than that of LM in practice.

\subsection{Acoustic model fusion by second pass rescoring}
\label{sec:fusion-practice}

In order to explicitly inject the pronunciation knowledge into the system, we further decompose the internal AM with approximation built on Markov and Viterbi assumption:
\begin{equation}
\begin{aligned}
P_{IAM}(X | \hat{Y}) & =  \sum_{\hat{P}}{P(\hat{P} | Y) P(X| Y, \hat{P})} \\
& \approx  \max_{\hat{P}}{P(\hat{P} | Y) P(X| \hat{P})}\\
\end{aligned}
\end{equation}
where $\hat{P}$ denotes a phoneme sequence, $P(\hat{P} | Y)$ represents an internal pronunciation model learned from the training data, and $P(X | \hat{P})$ represents an internal phoneme based AM. 
Conceptually, such phoneme-based AM is identical to the AM used in the hybrid ASR system \cite{dahl2011context}.
Theoretically, phonetic units are better trained as they are more evenly distributed in the training data. 

\begin{figure}[h]
\centering
\includegraphics[width=0.475\textwidth]{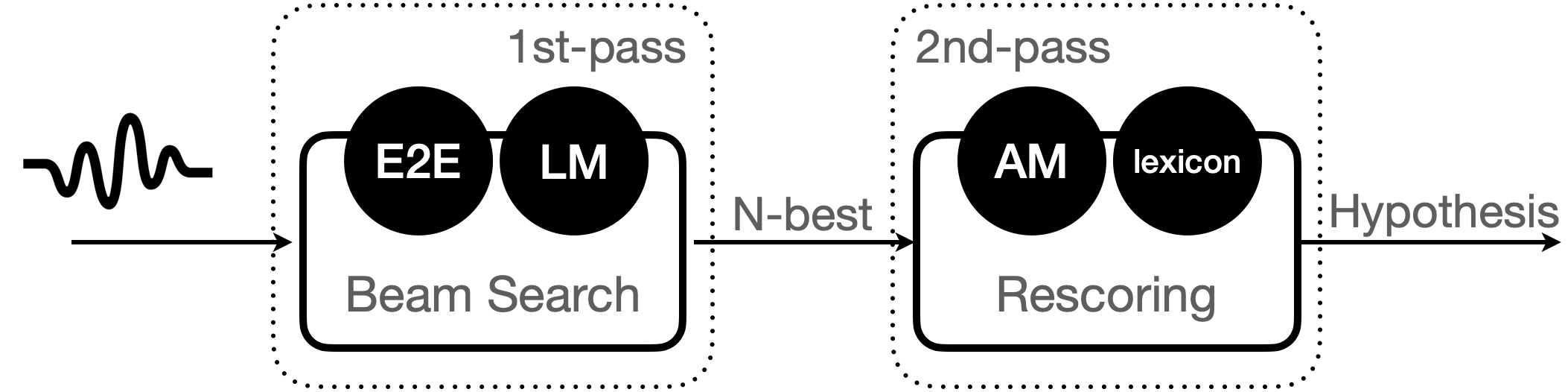}
\caption{AM fusion applied in the second pass ASR}
\label{fig:fusion}
\end{figure}

It becomes challenging when applying phoneme-based AM fusion and word-based LM fusion 
on a wordpiece \cite{kudo2018sentencepiece} E2E ASR model in a synchronous fashion during streaming, as the granularity of the modeling units of these three components are different. 
In practice, we separate the recognition as a two-pass process. 
The first pass operates in a streaming mode. 
We represent the E2E model and the external LM as WFSTs \cite{klambauer2017self} and formulate the first pass as a beam search of the composed WFSTs. 
The inverse of the ILM can be composed with the external LM to achieve ILM negation. 
The N-best hypotheses are preserved as the input to the second pass. 
In the second pass, an external AM score is computed by force-aligning the N-best hypotheses with the audio provided an lexicon.
By force alignment, the best possible phonetic sequence leading to the target word sequence is found at minimal cost. 
This cost is interpolated as Eq. (\ref{eq:amf}).
The system is depicted in Figure \ref{fig:fusion}.

\section{Experiments}
\label{sec:expts}

\subsection{Datasets and models}
\label{subsec:setup}

\begin{table*}[h]
    \centering
    \begin{tabular}{l|lll}
        \hline
         AM & modeling units & objective function & \#parameters \\
         \hline
         Conformer-wp & wordpiece & {CTC + AED}$^+$ & 120m$^{++}$\\
         Conformer-mono & monophone & CrossEntropy + MMI \cite{vesely2013sequence} & 13m\\
         SNDCNN-tri & triphone & CorssEntropy + MMI & 17m \\
         \hline
    \end{tabular}
    \caption{AM used through out the experiments. $^+$Though the model is trained with a multitask objective of CTC and AED, AED rescoring is not used during inference. $^{++}$Because AED is not included, the effective number of parameters during inference is 72m.}
    \label{tab:am}
\end{table*}

\subsubsection{Virtual assistant and dictation}
\label{sec:siri}

To evaluate AM fusion, especially for named entities, we created three datasets:
\begin{description}
    \setlength\itemsep{0.01em}
    \item[VA] The \textbf{V}irtual \textbf{A}ssistant dataset consists of 49k anonymized human-annotated virtual assistant queries. These queries often contain named entities and/or personal mentions.
    \item[DICT] The \textbf{DICT}ation dataset is made up of 23k anonymized human-annotated dictated sentences. 
    \item[TTS] The \textbf{T}ext-\textbf{T}o-\textbf{S}peech dataset have 103k synthesized audio-text pairs. We purposely inject named entities to every utterance. The TTS dataset examines ASR systems' ability of speech entity recognition. 
\end{description}

We trained a few AMs (\textit{Conformer-wp}, \textit{Conformer-mono}, and \textit{SNDCNN-tri}) on 50k-hour audio-text pairs with different modeling units, summarized in Table \ref{tab:am}.
Following \cite{DBLP:conf/interspeech/ZhangWPSY00YP022,dblp:conf/interspeech/YaoWWZYYPCXL21} and \cite{huang2020sndcnn}, we implemented Conformer and SNDCNN using the parameters recommended by those works, except that (1) the output sizes are adjusted to fit our in-house wordpiece/lexicon and (2) the number of conformer blocks of \textit{Conformer-mono} is reduced from 12 to 2 and the subsampling is done by filter bank concatenation followed by a fully-connected layer.
The conformer models are stacks of alternating transformer and pointwise/depthwise convolution while the SNDCNN models are modified resnet50 \cite{he2016deep}.
Though \textit{Conformer-wp} is trained with a multitask objective of CTC and AED, we carry out the benchmark in a streaming setup of a 240ms decoding window without AED rescoring.
The training data is a mixture curated in the same manner as VA and DICT.
A word-based 4-gram LM is trained on large amounts of in-domain text.
The LM is optimized for named entities so it greatly improves over VA and TTS testsets. Specifically, to improve recognition of personal content in the VA test set, the E2E+LM system is personalized using class-based language model and pronunciation-driven subword tokenization presented in \cite{lei2023personalization}.

We created a few baseline systems with and without an LM.
CTC prefix beam search is able to operate on \textit{Conformer-wp} without an external LM.
We built four AM fusion enhanced ASR systems by applying either \textit{SNDCNN-tri} or \textit{Conformer-mono} on top of \textit{Conformer-wp}.
Besides WER, we also measure oracle WER. 
The hypothesis closest to the reference in the N-best list is referred as oracle hypothesis.
Oracle WER represents the lowest achievable WER assuming the oracle hypothesis is chosen.

\begin{table*}[t]
    \centering
    \begin{tabular}{l|c|rl|lll|rrr}
    \hline
        & \multirow{2}{*}{id} & & \multirow{2}{*}{System} & \multicolumn{3}{|c|}{WER / Oracle} & \multicolumn{3}{c}{WERR \small{w/ AM fusion}}  \\
        \cline{5-10}
        & & & & \multicolumn{1}{c}{VA} & \multicolumn{1}{c}{DICT} & \multicolumn{1}{c|}{TTS} & VA & DICT & TTS \\
        \hline
        \multirow{4}{*}{baseline} & b2.1 & & SNDCNN-tri + LM & 3.99 & 3.68 & 12.6 & - & - & - \\
        & b2.2 & & Conformer-mono + LM & 13.77 & 14.13 & 21.3 & - & - & - \\
        & b2.3 & & Conformer-wp & 8.33 / 4.05 & 4.18 / 2.15 & 20.29 / 13.31 & - & - & - \\
        & b2.4 & & Conformer-wp + LM & 3.63 / 1.97 & 3.30 / 2.34 & 13.7 / 10.2 & - & - & - \\
        \hline
        \multirow{4}{*}{\shortstack[c]{AM\\fusion}} 
        & a2.1 & b2.3 + & SNDCNN-tri & 7.50 & 3.80 & 17.88 & 9.96\% & 9.09\% & 11.88\% \\
        & a2.2 & b2.3 + & Conformer-mono & 7.75 & 3.92 & 18.48 & 6.96\% & 6.22\% & 8.92\% \\
        & a2.3 & b2.4 + & SNDCNN-tri & 3.11 & 2.85 & 12.47 & 14.33\% & 13.64\% & 8.98\% \\
        & a2.4 & b2.4 + & Conformer-mono & 3.41 & 3.13 & 12.91 & 6.06\% & 5.15\% & 5.77\% \\
    \hline
    \end{tabular}
    \caption{WER, oracle WER and WERR of the baseline systems (first 4 systems) and the AM fusion-enhanced systems (the last 4 systems) on the VA, DICT and TTS datasets. LM is fused in first pass. AM is fused in second pass. Oracle of non-baseline systems is not shown.
 }
    \label{tab:siri-wer}
\end{table*}

\subsubsection{Librispeech}
\label{sec:lbs}


Using the same hyper-parameters as above, we trained another set of \textit{Confomer-wp} and  \textit{Confomer-mono} on the Librispeech \cite{panayotov2015librispeech} dataset. 
Librispeech is widely used in the community for ASR benchmark purposes.
\textit{Confomer-wp} is competitive with publicly reported numbers, yielding 2.5\% and 7.13\% WER on the clean test partition \textit{test-clean} and the noisy test partition \textit{test-other} respectively when full attention and AED rescoring are used \footnote{\cite{DBLP:conf/interspeech/ZhangWPSY00YP022} reports 2.85\% for test-clean and 7.24\% for test-other with the exact same architecture. The acoustic encoder uses full attention and the N-best is rescored by AED.}. 
However, for the rest of the analysis, we focus on streaming mode, decoding audio with 240ms chunk and CTC without AED rescoring.

\begin{table*}[h]
    \centering
    \begin{tabular}{l|c|rl|rr|rr}
        \hline
        & \multirow{2}{*}{id} & & \multirow{2}{*}{System} & \multicolumn{2}{c|}{WER} & \multicolumn{2}{c}{WERR \small{w/ LM/AM fusion}} \\
        \cline{5-8}
        & & & & test-clean & test-other & test-clean & test-other \\
        \hline
        \multirow{4}{*}{\shortstack[l]{baseline\\(in-domain)}} & b3.1 & &  Conformer-mono + LM & 8.51 & 19.12 & - & - \\
        & b3.2 & & Conformer-wp & 4.82 & 11.75 & - & - \\
        & b3.3 & & \ \ \ \ + LM (2nd-pass) & 4.70 & 11.26 & 2.49\% & 4.17\% \\
        & b3.4 & & \ \ \ \ + LM (1st-pass) & 3.95 & 9.33 & 18.05\% & 20.60\% \\
        \hline
        baseline (out-of-domain) & b3.5 & & Conformer$^{*}$-mono + LM & 18.86 & 39.56 & - & - \\
        \hline
        \multirow{2}{*}{\shortstack[l]{AM fusion\\(in-domain)}} & a3.1 & b3.2 + &Conformer-mono & 4.50 & 11.39 & 6.64\% &3.06\% \\ 
        & a3.2 & b3.4 + & Conformer-mono & 3.63 & 8.70 & 8.10\% & 6.75\% \\
        \hline
        \multirow{2}{*}{\shortstack[l]{AM fusion\\(out-of-domain)}} & a3.3 & b3.2 + & Conformer$^{*}$-mono & 4.54 & 11.47 & 5.81\% & \ \ 2.38\% \\
        & a3.4 & b3.4 + & Conformer$^{*}$-mono & 3.74 & 8.91 & 5.32\% & 4.50\% \\
        \hline
    \end{tabular}
    \caption{WER and WERR of Librispeech. $^{*}$The out-of-domain models are trained from data described in Section \ref{sec:siri}.}
    \label{tab:lbs-wer}
\end{table*}

\begin{figure}[h]
\centering
\includegraphics[width=0.475\textwidth]{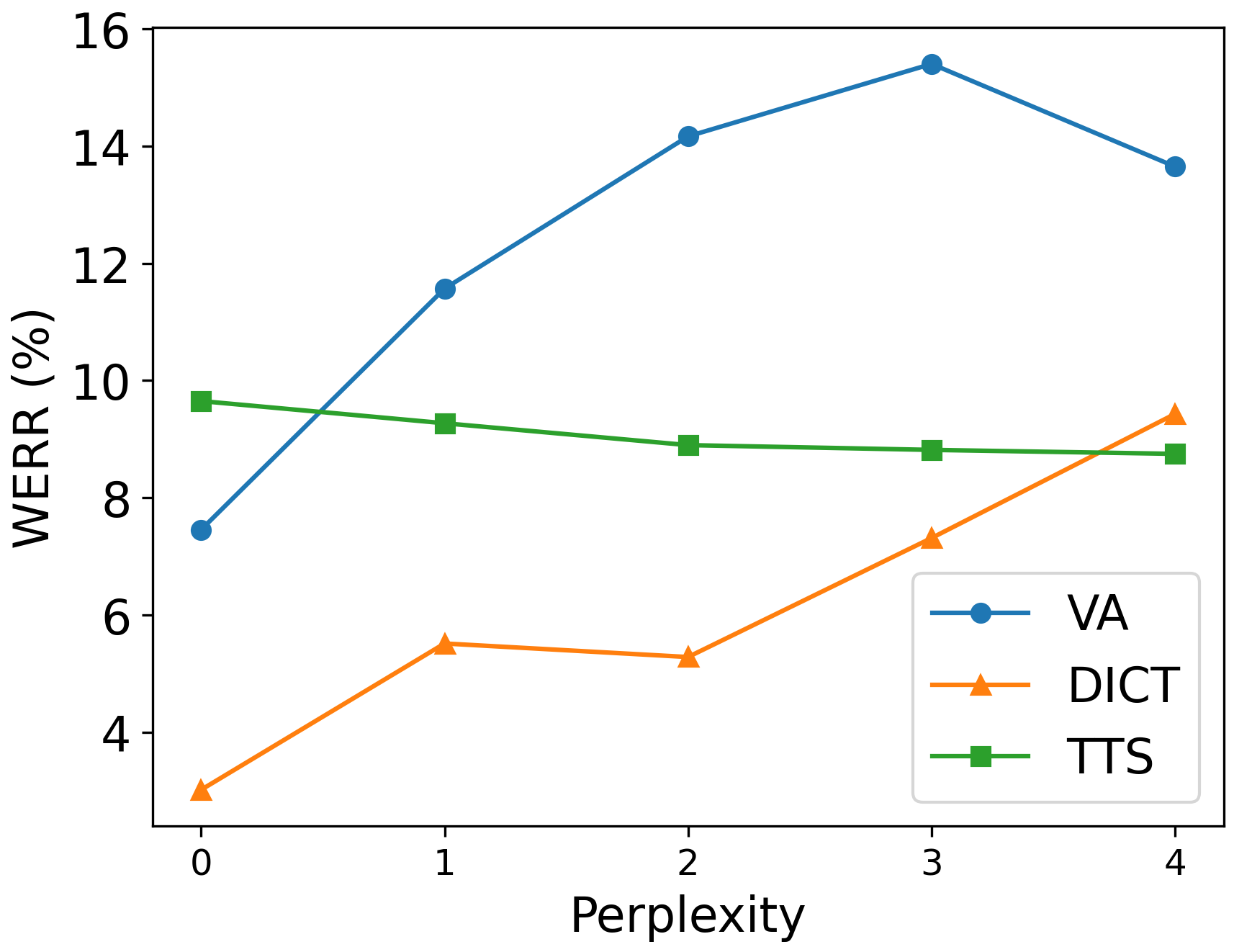}
\caption{WERR of different PPL buckets. The x-axis is bucket ID. Larger bucket ID corresponds to higher PPL.}
\label{fig:werr-vs-ppl}
\end{figure}

\subsection{Results and analysis}

Table \ref{tab:siri-wer} summarizes the WER, the Oracle WER and the WERR of the VA, DICT and TTS datasets. 
From the baseline, we can rank the modeling power of the AMs, i.e. \textit{Conformer-wp} $>$ \textit{SNDCNN-tri} $>$ \textit{Conformer-mono}.
We use the system ID in the tables to refer to our systems for the rest of the analysis, e.g. a2.4 refers to the AM fusion system that applies \textit{Conformer-mono} on the b2.4 baseline (\textit{Conformer-wp + LM}).
Because the VA and TTS datasets are entity-rich, and personalization \cite{lei2023personalization} is applied to boost recogntion of personal content in VA, \textit{Conformer-wp} with LM drastically outperforms its non-LM counterpart on VA and TTS while staying closer on DICT.
\textit{Conformer-mono} is notably weaker than the other AMs as its WER is one order of magnitude higher on VA and DICT even with the guidance of LM. 

\textit{Conformer-wp} represents the E2E ASR system.
As described in Section \ref{sec:fusion-practice}, for simplicity, our experiments incorporate AM fusion in the second pass.  
We first apply AM fusion on top of \textit{Conformer-wp} without LM using \textit{Conformer-mono}. 
At the first glance, the improvement is not as significant as LM fusion. 
However, this is expected because the oracle gap of b2.3 and b2.4 indicates that (1) LM is fused in the first pass such that it helps expand the search space and (2) AM is fused in the second pass, which is limited by the N-best quality.
We then apply AM fusion on top of LM fusion. The relative improvement using \textit{Conformer-mono} is similar to AM fusion in the non-LM setup.
WERR is larger on TTS and VA than on DICT, which shows that AM fusion helps under-trained words in named entities.
We then conduct AM fusion with \textit{SNDCNN-tri}. As indicated by b2.1 vs b2.2, \textit{SNDCNN-tri} is a more powerful AM than \textit{Conformer-mono}, leading to a significantly greater AM fusion WERR, especially when LM fusion is also employed. Despite this, \textit{Conformer-mono} is still of great interest because it uses the same front layers as \textit{Conformer-wp} so we can eventually combine it with \textit{Conformer-wp} through layer sharing and multitask training. By doing so, we can potentially achieve decent AM fusion improvement with little inference cost.

We next try to understand how AM fusion affects rare word recognition. 
We train a wordpiece N-gram LM on the E2E training data. 
We divide each dataset into five buckets of equal size based on wordpiece PPL.
Infrequent words fall into the high-PPL bucket with larger chance.
We compute each bucket's WERR and plot them in Figure \ref{fig:werr-vs-ppl}.
We can see that on \textit{VA} and \textit{DICT}, as PPL increases, higher WERR is achieved, which confirms the effectiveness of AM fusion in improving under-represented wordpieces. 
As we purposely inject named entities into every TTS utterance, the PPL difference between buckets is tiny, which explains why the TTS curve is flat.

We examine the effectiveness of AM fusion on Librispeech in a similar way. 
In these experiments, we only use \textit{Conformer-mono} as the representative of the phoneme-based models.
Additionally, to make a fair comparison between LM fusion and AM fusion, we build a new baseline where LM is used for second pass rescoring.
We can see that AM fusion provides more improvement than LM fusion (b3.3 vs a3.1) when they operate on the same search space.
No matter whether the base system is fused with LM or not, in-domain AM fusion brings consistent improvement.
We next take \textit{Conformer-mono} from the previous experiment. 
There exists clear domain mismatch between the AMs (comparing b3.1 and b3.5).
We fuse the out-of-domain phoneme-based model with the E2E ASR model with or without LM.
Because the AMs are not acoustically consistent, the out-of-domain \textit{Conformer-mono} contributes less improvement than the in-domain \textit{Conformer-mono} (e.g. a3.1 vs a3.3, a3.2 vs a3.4).
Though the out-of-domain \textit{Conformer-mono} has huge mismatch, the gain from out-of-domain \textit{Conformer-mono} AM fusion is still significant,
indicating that a large amount of improvement might come from the external knowledge injected by the explicit lexicon.

\section{Conclusion and future work}
\label{sec:conclusion}
In this work, we proposed AM fusion, which integrates an external AM into the E2E system via second pass rescoring. An AM score is computed for each first pass hypothesis in the N-best list for re-ranking. While second pass AM fusion brings significant improvement, it is limited by the first pass accuracy. Hybrid systems can work well on unusually spelled words, such as foreign names, but those words might not even appear in the N-Best list output by the E2E system in the first pass. Therefore, first pass AM fusion is appealing and can potentially help the recognition substantially. Our next milestone will be streaming AM fusion in the first pass.

\section{Acknowledgement}
We would like to thank Xiaoqiang Xiao, Christophe Van Gysel, Daben Liu, Ben Alexander and Russ Webb's helpful reviews and suggestions.

\bibliographystyle{IEEEbib}
\bibliography{refs}

\end{document}